\documentclass[11pt,letterpaper]{article}
\pdfoutput=1

\usepackage{jcappub}

\usepackage{afterpage}
\usepackage{capt-of}
\usepackage{diagbox}

\usepackage[figuresright]{rotating}
\usepackage{amsmath}
\usepackage{amssymb,amsmath,latexsym,graphics, graphicx,epsfig,multirow,comment,hyperref,appendix,feyn,slashed,xcolor,afterpage, makecell} 
\usepackage{booktabs}
\usepackage{tabularx}

\newcommand{\FeH}{\text{[Fe/H]} }

\newcommand {\be} {\begin {equation}}
\newcommand {\ee} {\end {equation}} 

\newcommand {\bes} {\begin {equation*}}
\newcommand {\ees} {\end {equation*}}

\newcolumntype{L}[1]{>{\raggedright\let\newline\\\arraybackslash\hspace{0pt}}m{#1}}
\newcolumntype{C}[1]{>{\centering\let\newline\\\arraybackslash\hspace{0pt}}m{#1}}
\newcolumntype{R}[1]{>{\raggedleft\let\newline\\\arraybackslash\hspace{0pt}}m{#1}}

\usepackage{csvsimple}

\makeatletter
\newcommand\footnoteref[1]{\protected@xdef\@thefnmark{\ref{#1}}\@footnotemark}
\makeatother

\DeclareRobustCommand{\Sec}[1]{Sec.~\ref{#1}}

\DeclareRobustCommand{\Tab}[1]{Table~\ref{#1}}

\DeclareRobustCommand{\Fig}[1]{Fig.~\ref{#1}}

\DeclareRobustCommand{\Eq}[1]{Eq.~(\ref{#1})}

\newcommand{\beq}{\begin{equation}}
\newcommand{\eeq}{\end{equation}}

\begin{document}

\title{
The Metal-Poor Stellar Halo in RAVE-TGAS and its Implications \\ \vspace{0.05in} 
for the Velocity Distribution of Dark Matter
}

\author[a]{Jonah Herzog-Arbeitman,}
\author[a]{Mariangela Lisanti,}
\affiliation[a]{Department of Physics, Princeton University, Princeton, NJ 08544}

\author[b]{Lina Necib}
\affiliation[b]{Center for Theoretical Physics, Massachusetts Institute of Technology, Cambridge, MA 02139}

\emailAdd{jonahh@princeton.edu}
\emailAdd{mlisanti@princeton.edu}
\emailAdd{lnecib@mit.edu}

\date{today}

\abstract{
The local velocity distribution of dark matter plays an integral role in interpreting the results from direct detection experiments.  We previously showed that metal-poor halo stars serve as excellent tracers of the virialized dark matter velocity distribution using a high-resolution hydrodynamic  simulation of a Milky Way--like halo.  In this paper, we take advantage of the first \textit{Gaia} data release, coupled with spectroscopic measurements from the RAdial Velocity Experiment (RAVE), to study the kinematics of stars belonging to the metal-poor halo within an average distance of $\sim$$5$~kpc of the Sun.  We study stars with iron abundances $\FeH < -1.5$ and $-1.8$ that are located more than 1.5~kpc from the Galactic plane.  Using a Gaussian mixture model analysis, we identify the stars that belong to the halo population, as well as some kinematic outliers.  We find that both metallicity samples have similar velocity distributions for the halo component, within uncertainties.  Assuming that the stellar halo velocities adequately trace the virialized dark matter, we study the implications for direct detection experiments.  The Standard Halo Model, which is typically assumed for dark matter, is discrepant with the empirical distribution by $\sim$$6\sigma$, predicts fewer high-speed particles, and is anisotropic.  As a result, the Standard Halo Model overpredicts the nuclear scattering rate for dark matter masses below $\sim$$10$~GeV.  The kinematic outliers that we identify may potentially be correlated with dark matter substructure, though further study is needed to establish this correspondence.  
}

\preprint{MIT-CTP/4928}
\maketitle
%\pagebreak

\section{Introduction} 
\label{sec:intro}

Little is known about the velocity distribution of dark matter~(DM) in the Solar neighborhood, and yet the details impact its detectability.  For example, the rate of DM traversing the Earth and scattering off a target depends on both its velocity, as well as its interaction cross section with the material~\citep{Goodman:1984dc,Drukier:1986tm}.  Direct detection experiments, which are built in low-background underground laboratories, aim to discover these weakly interacting particles.  However, the interpretation of their results are dependent on the uncertainties associated with the local DM phase space~\citep{Freese:2012xd,Cremonesi:2013bma,Green:2017odb}.  In this paper, we discuss how such uncertainties can be mitigated by empirically determining the velocity distribution of DM with astrometric measurements of the metal-poor stars in the stellar halo.

There is currently no universal form of the DM velocity distribution that is consistent with results from numerical simulations.  The standard is to simply assume that the DM velocities follow a Maxwell-Boltzmann distribution with a peak velocity of $\sim$$220$~km/s.  Referred to as the Standard Halo Model (SHM), this distribution is derived from the assumption that the DM halo is well-modeled as a collisionless gas; an isothermal distribution with mass density $\rho \sim 1/R^2$, as naively expected due to the flat rotation curve, corresponds to a Maxwellian velocity distribution \citep{Drukier:1986tm}.  Additional forms have also been suggested~\citep{Hansen:2004qs,Chaudhury:2010hj, Lisanti:2010qx, 2012JCAP...05..005C, 2013JCAP...12..050B,Fornasa:2013iaa}, though they typically rely on assumptions such as equilibrium and/or isotropy of the local DM distribution.  

A separate approach has been to take the velocity distribution directly from $N$-body simulations, which trace the multi-body gravitational interactions in a galaxy's hierarchical formation process.  DM-only simulations typically find deviations from the SHM, with distributions that extend to higher velocities and show a deficit in the peak region~\citep{Vogelsberger:2008qb,MarchRussell:2008dy,Fairbairn:2008gz,Kuhlen:2009vh,Mao:2012hf}. Some of this tension appears to be relieved when baryonic physics is included in the simulation~\citep{Ling:2009eh,Pillepich:2014784,Bozorgnia:2016ogo, Kelso:2016qqj, Sloane:2016kyi, Bozorgnia:2017brl}. These results are sensitive to the different realizations of Milky Way--like halos in the simulations, which emphasizes the importance of obtaining the true velocity distribution empirically.

The challenges associated with modeling the DM's velocity distribution have led to sustained efforts to establish astrophysics-independent analysis strategies for direct detection experiments~\citep{Fox:2010bz,Fox:2010bu,Frandsen:2011gi,Gondolo:2012rs,HerreroGarcia:2012fu,DelNobile:2013cva,Feldstein:2014gza,Fox:2014kua, Feldstein:2014ufa,Anderson:2015xaa,Blennow:2015gta,Herrero-Garcia:2015kga,Gelmini:2015voa,Gelmini:2016pei,Gelmini:2017aqe, Gondolo:2017jro,Ibarra:2017mzt,Fowlie:2017ufs}.   These methods are primarily useful in comparing the consistency of results between different experiments.  However, one still needs to know the velocity distribution to accurately infer the bounds on (or observed properties of) the DM mass and scattering cross section.

Previously, we showed that local metal-poor halo stars serve as effective kinematic tracers of the virialized DM using the \textsc{Eris} cosmological simulation~\citep{eris_paper}.  %We demonstrated the correspondence between the kinematics of the DM and metal-poor stars explicitly.  
\textsc{Eris} is one of the highest-resolution hydrodynamic simulations of a Milky Way--like galaxy~\citep{Guedes:2011ux,Guedes:2012gy,Pillepich:2014784,Pillepich:2014jfa,Genel:2014lma,2017MNRAS.469.4012S,2013ApJ...763...65A}.  The DM and stellar velocity distributions in \textsc{Eris} approach each other as more metal-poor stars are considered.  The results suggest that the velocity distribution of the local DM can be obtained directly from the corresponding distribution of the metal-poor stellar halo.  Further study is needed using different $N$-body simulations to understand how the results drawn from \textsc{Eris} generalize to different merger histories, but these initial findings suggest a promising new method to trace the local DM velocities.  

The proposal in~\cite{eris_paper} makes no assumptions regarding the equilibrium of the stellar halo.  This is fundamentally different than the alternate possibility of mapping the gravitational potential with stellar tracers and inferring the velocity distribution using Jeans theorem.  Because our approach makes no assumption about whether the halo is in dynamic steady-state, it is particularly well-suited for characterizing the local DM phase space, which may exhibit substructure due to nearby mergers.  

The reason that metal-poor stars serve as kinematic tracers for DM is that both primarily originate from satellites that merged with the Milky Way~\citep{1978MNRAS.183..341W, Johnston:1996sb, Helmi:1999uj, Helmi:1999ks, Bullock:2000qf, Bullock:2005pi,Robertson:2005gv,Font:2005qs,Font:2005rm,Cooper:2009kx}.  %To probe this old stellar population, one can sample stars with low iron abundance, [Fe/H]. A star's metallicity provides hints to its origin.  In particular, those that originated from low-mass satellites typically exhibit lower metallicities because star formation in these systems turns off when gas is stripped during tidal disruption.   
In \textsc{Eris}, the convergence in the kinematic distributions improves as one pushes towards very metal-poor stars ($\FeH \lesssim -3$).  This likely arises from the fact that both the smallest ultrafaint dwarf galaxies as well as the larger classical dwarf galaxies contribute stars in this range~\citep{Deason:2016wld}; while more massive dwarfs are typically more metal-rich, on average, than the ultrafaints, the metallicity distribution of their stellar constituents can extend down to very low values~\citep{Kirby:2008ab}.  As a result, probing the kinematics of the most metal-poor stars samples a broad mass range for the disrupted satellites in a galaxy's merger history.  Because the dark matter halo is built from all these mergers, it is well-modeled by the most metal-poor stars in the Galaxy.

In this paper, we present the first dedicated study of the stellar halo kinematics with the goal of reconstructing the local DM velocity distribution, which is relevant for direct detection experiments.  We emphasize that the correlation between the kinematics of DM and metal-poor stars discussed in \cite{eris_paper} and assumed in this work is based entirely on results from numerical simulations, although it follows from the picture of hierarchical formation of halos.  We take advantage of the first data release (DR1) from \emph{Gaia} to characterize the kinematics of the metal-poor halo within $\sim$5~kpc of the Sun and more than 1.5~kpc from the Galactic plane~\citep{2016A&A...595A...2G,2016A&A...595A...4L}.  The \emph{Tycho}-\emph{Gaia} astrometric solution (TGAS)~\citep{2015A&A...574A.115M} provides the full proper motions for the subset of stars with overlap in the \textsc{Hipparcos} and \emph{Tycho}-2 catalogs~\citep{1997ESASP1200.....E, 2000A&A...355L..27H,vanLeeuwen:2007tv}.  Combining these results with line-of-sight velocities from the RAdial Velocity Experiment (RAVE)~\citep{2017AJ....153...75K} yields one of the most detailed maps of the six-dimensional phase space of the local stellar halo.  
Although the uncertainties on the parallaxes and proper motions of the stars in the RAVE-TGAS (hereafter referred to as RT) catalog have not reached their projected values for the mission's end, we can develop an analysis procedure that can be further refined as the data improve.  

Using the kinematic distribution of the metal-poor halo to infer the DM properties, we find that the DM velocities are slower than typically assumed, weakening direct detection limits on nuclear scattering for DM masses below  $\sim$$10$~GeV. We also identify kinematic outliers in the stellar population, which could potentially be associated with DM substructure, though further study is required to confirm their origin.  

This paper is organized as follows. In \Sec{sec:data}, we describe the RT catalog as well as the selection cuts used to extract the local metal-poor halo stars.  \Sec{sec:stats} describes the mixture model used to identify the halo stars and kinematic outliers, and \Sec{sec:kinematicdistribution} presents the results of applying the study on data.   \Sec{sec:DM} explores the implications of the recovered velocity distributions for direct detection experiments.  We conclude in \Sec{sec:conclusions} and provide extended results in the Appendix.

\section{Catalogs and Selection} 
\label{sec:data} 

\subsection{Stellar Catalogs}
 Launched in 2013, \emph{Gaia} has a nominal mission lifetime of five years and will ultimately achieve unprecedented accuracies on the positions and proper motions of stars brighter than $V\approx20$~mag.  DR1, which includes results from the first 14 months of the mission, contains the positions of more than a billion stars, as well as their mean G-band fluxes and magnitudes \citep{2016A&A...595A...2G, 2016A&A...595A...4L}.  It is possible to extract the proper motions and parallaxes for the $\sim$2~million of these stars that have a match in the \textsc{Hipparcos} and \emph{Tycho}-2 catalogs.  These stars, which comprise the TGAS catalog, have typical accuracies of $\sim$0.3~mas on position and 1~mas/year on proper motion \citep{2015A&A...574A.115M}.

Combining the TGAS catalog with the RAVE fifth data release (DR5) provides a unique opportunity to study the local phase space of the stellar halo~\citep{2017AJ....153...75K}.  RAVE\footnote{\url{https://www.rave-survey.org/project/}} is a magnitude-limited ($9 < I < 12$) catalog of stars from the Southern hemisphere~\citep{2011AJ....142..193B,2017AJ....153...75K}.  DR5 includes 457,588 unique stars observed from 2003--2013. In addition to the magnitude selection, the RAVE stars within the Galactic plane ($|b| < 25^\circ$) satisfy the color requirement $J-K_s \geq 0.5$, imposed to bias the survey towards giants.  All stars satisfying these magnitude and color criteria were randomly sampled. While the catalog is not volume-complete, it exhibits no kinematic or chemical biases, which is critical for the work in this paper~\citep{2017MNRAS.468.3368W}.  A wealth of information is provided for the 255,922 common stars between RAVE and TGAS, including radial velocities, proper motions, chemical abundances, and stellar parameters. The combined data set provides the most detailed snapshot of the kinematics of the local stellar distribution to date.

For the iron abundances [Fe/H] of the stars, we use the RAVE-on catalog~\citep{2016arXiv160902914C}.  This catalog was built using The Cannon~\citep{2015ApJ...808...16N} to perform a data-driven re-analysis of the RAVE spectra.  The typical uncertainty on elemental abundances in RAVE-on is $\sim$0.07~dex, which is more precise than the $\sim$0.2~dex errors on [Fe/H] derived from the RAVE chemical pipeline~\citep{2017AJ....153...75K}. 

Additionally, we use the stellar distances as provided by~\cite{2017arXiv170704554M}.  These distances are derived by combining the \emph{Gaia} parallaxes and updated measurements of $T_\text{eff}$ using the Infrared Flux Method~\citep{1977MNRAS.180..177B,1979MNRAS.188..847B} with the RAVE Bayesian distance estimator~\citep{Burnett:2010jy,Binney:2013ppa}.
The combination of data sets reduces the uncertainty on the distances by nearly a factor of two from the spectrophotometric values provided in RAVE DR5.  

In summary, we take the proper motions and their associated errors, as well as the galactic $(l,b)$ and equatorial  $(\alpha, \delta)$ coordinates, from the TGAS catalog.  The heliocentric radial velocities plus their associated errors  are taken from the RAVE DR5 catalog.  The iron abundance [Fe/H] comes from the RAVE-on catalog, and the distance estimates from~\cite{2017arXiv170704554M}.

\subsection{Selection Cuts}
\label{sec:selection}

To select stars with the best radial velocities and stellar parameters, we apply the following quality cuts to the data:  $\sigma$(HRV) $\leq$ 8~km/s, $|$\texttt{CorrectionRV}$|$ $\leq$ 10~km/s, \texttt{correlationCoeff} $\geq$ 10, SNR $\geq$ 20, \texttt{AlgoConv} $\neq$ 1, and \texttt{flag\_N}$\neq$1.   
The first three selection criteria select stars with well-measured radial velocities.  The signal-to-noise ratio (SNR) requirement selects stars where the stellar parameters are determined with high-confidence.  \texttt{AlgoConv} is a quality flag for the RAVE stellar parameter pipeline; if it is equal to unity, then the output of the pipeline is not reliable.  The requirement on \texttt{flag\_N} removes stars that are labeled as peculiar based on their spectral morphology~\citep{2012ApJS..200...14M}. We also remove duplicate stars with the same right ascension and declination\footnote{Since the velocities and chemical properties of the duplicate stars are within their associated uncertainties, for reproducibility, we list the IDs of the stars that we used in our analysis: $['20070918\_0235\rm{m}32\_042'$,  $'20060327\_1514\rm{m}16\_128'$,  $'20051012\_1944\rm{m}48\_091'$,  $'20060818\_0024\rm{m}21\_005'$,  $'20070911\_0410\rm{m}44\_090'$,  $'20060817\_2141\rm{m}51\_106']$.}.  
These quality cuts are applied to all samples analyzed in this paper. 

We caution the reader that the vast majority of the stars in our sample have $\log g\lesssim 2.0$ where the RAVE parallaxes have been observed to be systematically below their TGAS values.\footnote{These stars are labeled with \texttt{flag\_low\_logg}=1 in the catalog.}  This, in turn, could cause the distances---and thus the velocities---of the stars in our sample to be overestimated.  This systematic effect has not been accounted for in the distance estimates provided by~\cite{2017arXiv170704554M}
and requires further study.

\begin{figure*}[tb] %  figure placement: here, top, bottom, or page
   \centering
   \includegraphics[width=3.0in]{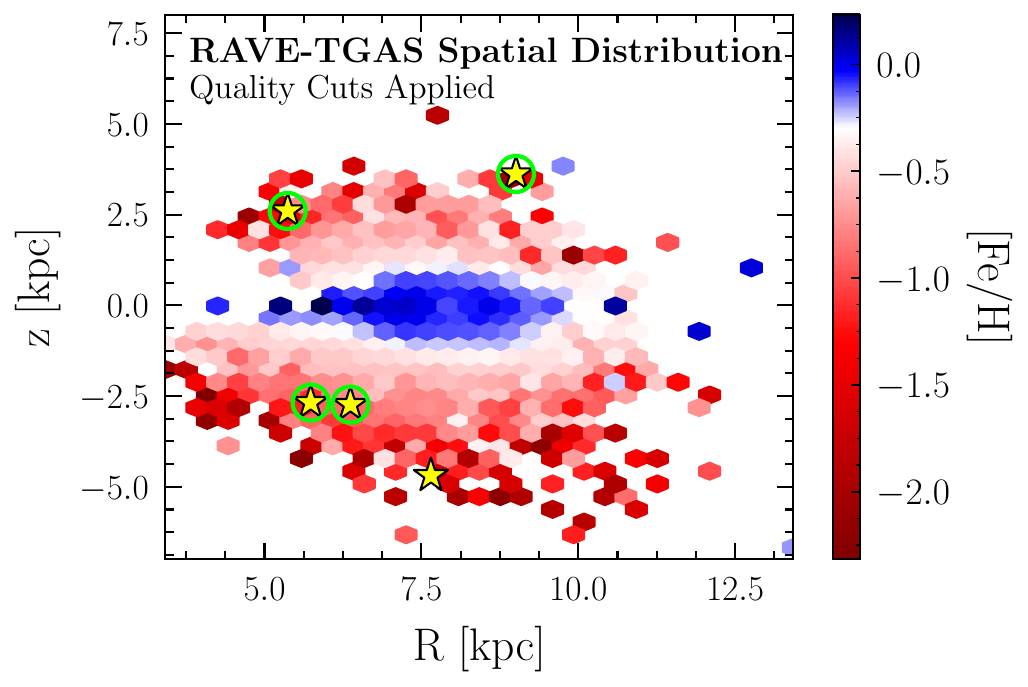} 
   \qquad
    \includegraphics[width=2.5in,trim={0 -0.0cm 0 0}]{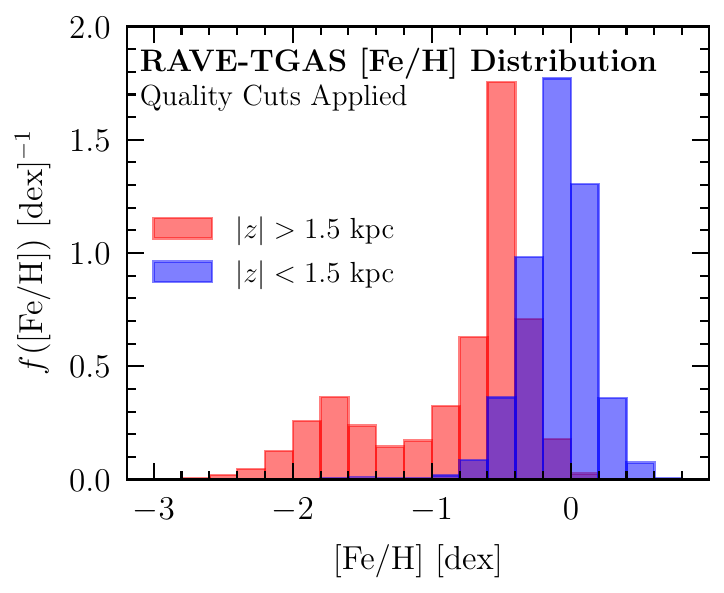}
   \caption{(Left) Distribution of the RAVE-TGAS sample in vertical distance from the Galactic plane, $z$, and galactocentric radius, $R$.  The average iron abundance, $\FeH$, of stars at each point is shown.  The yellow stars~(green circles) indicate kinematic outliers identified in the $\FeH <-1.5$~($-1.8$) sample.  See \Sec{sec:kinematicdistribution} for more details.  (Right)  Distribution of iron abundances for the RAVE-TGAS sample, as a function of $|z|$.  The low-metallicity halo component becomes more apparent farther from the plane.  }
   \label{fig:zRgc}
\end{figure*}

Figure~\ref{fig:zRgc} demonstrates some basic properties of the RT sample after quality cuts.  The left panel shows the distribution of iron abundance [Fe/H] as a function of distance off the Galactic plane, $z$, and galactocentric distance, $R$.  Note that the metallicities of all stars at a given point are averaged together in the figure.  The stars in the sample extend up to $\sim5$~kpc from the plane and are clustered within $\sim4$--12~kpc of the Galactic Center.  The average iron abundance of the stars decreases further from the plane, as one moves from the thin/thick disk-dominated regime to the halo regime.  

The right panel of Fig.~\ref{fig:zRgc} shows the distribution of iron abundances for stars above and below a vertical displacement of 1.5~kpc.  Throughout, a distance cut $|z| > z_0$, means that there is a $95\%$ probability that the measurement of $|z|$ passes the $|z| > z_0$ requirement.  This allows us to account for the large distance errors that propagate to the magnitude of the vertical displacement.  In obtaining the vertical coordinate of the stars, we assume that the Sun is located 15 pc above the Galactic mid-plane \citep{2001ApJ...553..184C,2001AJ....121.2737M,Juric:2005zr,2014ApJ...797...53G}.  When $|z| < 1.5$~kpc, the distribution of iron abundance is clearly peaked at $\FeH\sim 0$.  As one moves further above the disk plane, the distribution shifts to lower metallicities.  In particular, we see a broad peak appear at $\FeH\sim-1.7$, which is consistent with the local stellar halo population---see \emph{e.g.},~\cite{Ivezic:2008wk,Carollo:2007xh}.

To calculate the scattering rate in direct detection experiments, one needs to know the DM velocities in the laboratory frame.  Ideally one would map the velocity distribution of the metal-poor stellar halo as close to the Sun as possible.  Moving closer to the Galactic plane, however, increases the contamination from disk and \emph{in-situ} halo stars~\citep{Bond:2009mh,2010ApJ...712..692C}. For the extrapolation to DM, we are interested specifically in \emph{ex-situ} halo stars, which originate from mergers, unlike \emph{in-situ} halo stars, which are born in our Galaxy from \emph{e.g.}, gas condensation within the host and are thus more abundant closer to the plane~\citep{2009ApJ...702.1058Z,2011MNRAS.416.2802F,2012MNRAS.420.2245M,Pillepich:2014jfa,2015MNRAS.454.3185C,2017arXiv170405463B}.  A careful balance must therefore be achieved when defining the region of interest to maximize the probability of selecting stars that likely originate from Milky Way satellites and thus should be correlated with the DM.  Therefore, in addition to the quality cuts discussed above, we also place additional selection requirements on the vertical displacement and iron abundance of the stars. 

To reduce contamination of disk and \emph{in-situ} halo stars, we require that the stars satisfy $|z| > 1.5$~kpc.  In addition, we focus on the most metal-poor stars in the RT catalog.  The two benchmark scenarios that we consider have $\FeH<-1.5$ and $-1.8$. The number of stars remaining after these selection cuts is 141 and 69, respectively.  The stars in the $\FeH <-1.5$~($-1.8$) sample, range from 1.9--7~(2.7--7)~kpc from the Sun, with a mean distance of 4.5~(4.7)~kpc.  These selection criteria are meant to identify metal-poor stars that most likely belong to the halo.  We rely solely on chemical and spatial cuts to avoid potentially biasing the recovered velocity distribution through any kinematic cuts.   In Sec.~4, we describe the halo-like kinematic properties of the stars in our sample. In principle, the sample size can be improved by loosening the requirement on the vertical displacement, though one would likely need to adopt more sophisticated algorithms, such as extreme deconvolution~\citep{2011AnApS...5.1657B}, to distinguish the halo and disk contributions---see also~\cite{2017ApJ...835...81B}. 

Our cuts on $|z|$ and $\text{[Fe/H]}$ are meant to select \emph{ex-situ} stars, which are brought into the Milky Way through mergers.  \cite{2017arXiv170405463B} finds evidence for a metal-rich halo component in the RT catalog with $\FeH\sim-1.1$.  They argue that this component is likely comprised of \emph{in-situ} stars formed within the Milky Way. As our goal here is to infer the DM velocities, we are only interested in \emph{ex-situ} halo stars, which share a common origin with the DM in merging satellites.  For this reason, we focus on  stars with iron abundances below $-$1.5 and $-$1.8. Based on our studies of the \textsc{Eris} simulation~\citep{eris_paper}, one would ultimately want to consider stars with even lower metallicities to feel confident in the extrapolation to DM. We discuss the convergence of our study in \Sec{sec:DM}.  

\section{Statistical Methods} 
\label{sec:stats}

We model the velocities of the metal-poor RT sample\footnote{We use `metal-poor sample' as a shorthand for the sample selected as described in \Sec{sec:selection} using metallicity, vertical distance ($|z|>$ 1.5 kpc), and quality criteria cuts.} as a mixture of two multivariate normal distributions, each defined by 
\begin{equation}
\mathcal{N}(\mathbf{v}| \boldsymbol{\mu}, \mathbf{\Sigma} ) = \frac{\exp\left[ -\frac{1}{2} \left( \mathbf{v} -\boldsymbol{\mu} \right)^T \mathbf{\Sigma}^{-1} \left( \mathbf{v} - \boldsymbol{\mu} \right) \right] }{\sqrt{\left(2 \pi \right)^3 \text{det}\left(\mathbf{\Sigma}\right)}}\, ,
\label{eq:multivariate}
\end{equation}
where $\boldsymbol{\mu}$ is the vector of means.  In spherical coordinates, the covariance matrix $\mathbf{\Sigma}$ is defined as
\begin{align} \label{eq:covariance}
\mathbf{\Sigma} &=
  \left( {\begin{array}{ccc}
   \sigma_r^2 &\rho_{r \theta}  \sigma_r \sigma_\theta & \rho_{r \phi} \sigma_r \sigma_\phi  \\
   \rho_{r\theta} \sigma_r \sigma_\theta  & \sigma_\theta^2 &  \rho_{\theta \phi}\sigma_\theta \sigma_\phi \\
    \rho_{r\phi} \sigma_r \sigma_\phi &  \rho_{\theta \phi} \sigma_\theta \sigma_\phi  & \sigma_\phi^2 \\
  \end{array} } \right) \, ,
 \end{align}
where the $\sigma$ and $\rho$'s are the dispersions and correlation coefficients, respectively. The two distributions are intended to model the halo stars in the region-of-interest, as well as the population of kinematic outliers.  The stars in the RT sample are concentrated within $\sim$5~kpc of the Sun, with $|z|>1.5$~kpc.  The Galactocentric radial distances of the stars range from 4.1--11~kpc, with mean $7.5$~kpc, for both metallicity samples, suggesting that they are dominated by inner-halo stars.  At this stage, we remain agnostic to the origin of the outliers.  Our primary goal is to separate out the population of stars with kinematics distinct from the dominant halo population in the event that it is associated with phase-space substructure.  The procedure we adopt is a generalization of the statistical methods outlined in~\cite{2010arXiv1008.4686H, foreman_mackey_2014_15856} for outlier identification.       

The data set of interest consists of $N$ stellar velocities $\mathbf{v}_i$, where $i$ labels an individual star.  We use the public code \texttt{gala}~\citep{Price-Whelan:2017} to perform the coordinate transformations from the ICRS frame to the spherical Galactocentric frame, with the assumption that the Sun is located at a distance of 8~kpc from the Galactic center and a vertical distance $z=15$~pc above the plane.
The errors on $\mathbf{v}_i$ are calculated by propagating the stated uncertainties on line-of-sight velocities and proper motions, including the correlations between proper motions in right ascension and declination, assuming that they are Gaussian distributed~\citep{Bovy:2011aa}.\footnote{We ignore the uncertainties on the stellar positions in the error propagation, as they are subdominant to the uncertainties on distance and proper motion.}  Following this prescription, we obtain measurement errors $\boldsymbol{\sigma}_i$ for each star.  A given halo star is therefore drawn from a multivariate normal  where each component of the dispersion is modeled as 
\begin{equation}\label{eq:dispersions}
\sigma_{\text{h},i} = \sqrt{(\sigma_\text{h})^2 + (\sigma_i)^2} \, , 
\end{equation}
which convolves the true dispersion of the halo distribution $\sigma_\text{h}$ with the measurement error $\sigma_i$ of the $i^\text{th}$ star.  (Note that the unbolded $\sigma_i$ represents a particular component of $\boldsymbol{\sigma}_i$ in this shorthand.). The dispersions from \Eq{eq:dispersions}   are substituted into \Eq{eq:covariance}, which we hereafter label as the covariance matrix $\mathbf{\Sigma}_{\text{h}, i}$ for star $i$.  The same is done for the kinematic outliers, whose covariance matrix is labeled as $\mathbf{\Sigma}_{\text{ko}, i}$.

\begin{table}[t]
\begin{center}
\begin{tabular}{C{2cm}C{1.2cm}C{1.8cm}C{1.8cm}}
 %\multicolumn{3}{c}{\textsc{Priors}} \\
\Xhline{3\arrayrulewidth}
\renewcommand{\arraystretch}{1}
Parameter & Type&\multicolumn{2}{c}{Priors} \\
&& Halo & Outlier \\ 
\hline
$\mu_r$, $\mu_{\theta}$& linear&  $[-50,50]$ & $[-50,50]$\\
$\mu_{\phi}$& linear& $[-50,50]$ & $[-400,400]$ \\
$\sigma_r$& log &$ [10,300]$ & $[10, 1200]$ \\
$\sigma_{\theta}$& log& $[10, 300]$ & $[100, 4000]$ \\
$\sigma_{\phi}$& log & $ [5, 300] $ & $[60, 4000]$ \\
$\rho_{r \theta}$, $\rho_{r \phi}$, $\rho_{\theta \phi}$& linear & $[-1,1] $ & $[-1,1]$ \\
$Q$& log & --- & $[10^{-3}, 1]$  \\
\Xhline{3\arrayrulewidth}
\end{tabular}
\caption{\label{tab:priors} Parameters and associated prior types/ranges for the halo and outlier populations.}
\end{center}
\end{table}

A flag $q_i$ is assigned to each star, which labels whether or not it belongs to the halo---\emph{e.g.}, $q_i =0$(1) for halo(outlier) stars.  The likelihood for a halo star is 
\begin{equation}
p\left( \mathbf{v}_i \,|\, \boldsymbol{\sigma}_i, q_i = 0,\theta \right) = \mathcal{N} \left( \mathbf{v}_i \,|\, \boldsymbol{\mu}_\text{h}, \boldsymbol{\Sigma}_{\text{h},i} \right) \, ,
\end{equation}
where the model parameters $\theta \supset \left( \boldsymbol{\mu}_\text{h}, \boldsymbol{\Sigma}_\text{h} \right)$ are the true mean and covariance of the halo distribution.  Similarly, the likelihood for the kinematic outliers is 
\begin{equation}
p\left( \mathbf{v}_i \,|\, \boldsymbol{\sigma}_i, q_i = 1,\theta\right) = \mathcal{N} \left( \mathbf{v}_i \,|\, \boldsymbol{\mu}_\text{ko}, \boldsymbol{\Sigma}_{\text{ko},i} \right) \, ,
\end{equation}
with parameters $\theta \supset \left( \boldsymbol{\mu}_\text{ko}, \boldsymbol{\Sigma}_\text{ ko}\right)$. 
It follows that the total likelihood is 
\begin{equation}
p\left(\{\mathbf{v}_i\}  \,|\,\{\boldsymbol{\sigma}_i\}, \{q_i\}, \theta  \right) =  \prod_{i=1}^N \, p \left( \mathbf{v}_i \,|\, \boldsymbol{\sigma}_i, q_i, \theta \right)\, ,
\label{eq:unmarginalized}
\end{equation}
where the bracketed quantities are meant to denote the full list of $N$ elements.   The assumption that the kinematic outliers follow a normal distribution is not very important; indeed, we let the dispersions for the outliers vary over a wide range precisely to capture any stars whose velocities differ from the halo component.  

 At this stage, the likelihood carries a large number of free parameters, including the 18 Gaussian parameters in $\theta = ( \boldsymbol{\mu}_\text{h}, \boldsymbol{\Sigma}_\text{h}, \boldsymbol{\mu}_\text{ko}, \boldsymbol{\Sigma}_\text{ko})$ and the $N$ values of $q_i$.  Marginalizing over the $q_i$ reduces the likelihood to a mixture of two Gaussians:
%\begin{widetext}
\begin{eqnarray}
p\left(\{\mathbf{v}_i\}  \,|\,\{\boldsymbol{\sigma}_i\}, \theta \right) %&=&  \sum_{\{q_i\}} \, \prod_{i=1}^N \, p \left( \mathbf{v}_i \,|\, \boldsymbol{\sigma}_i, q_i, \theta \right) \\ \nonumber
&=&  \prod_{i=1}^N \,  \Big[ (1-Q) \, \, p \left( \mathbf{v}_i \,|\, \boldsymbol{\sigma}_i,  q_i = 0, \theta \right)  + Q  \, p\left( \mathbf{v}_i \,|\, \boldsymbol{\sigma}_i, q_i = 1, \theta \right) \Big] \, ,
\label{eq:marginalized}
\end{eqnarray}
%\end{widetext}
where $Q$ is the probability that a star is an outlier.  The posterior distributions for these parameters are obtained using the Markov Chain Monte Carlo \texttt{emcee} \citep{2013PASP..125..306F}.  For the results presented in this paper, we take the number of walkers to be 1200 and the number of samples per walker to be  800. 
 
The priors are listed in \Tab{tab:priors}.  Log priors are assumed for the dispersions as well as for $Q$, while all other parameters have linear priors.  We take much wider priors on the dispersions of the outlier population, as compared to the halo population, to reflect the assumption that the outliers have higher velocities.  In addition, the prior range on $\mu_\phi$ for the outliers is larger than the others to absorb potential contamination from disk stars.

We conclude by noting that our statistical procedure models the stellar halo as a single population, despite the fact that current evidence suggests that it is comprised of two separate components that differ in their spatial and velocity distributions, as well as their chemical abundances~\citep{Carollo:2007xh,2010ApJ...712..692C,2010ApJ...714..663D,2012MNRAS.422.2116K,2013ApJ...763L..17H,2013MNRAS.430.2973K,2013ApJ...763...65A,2015ApJ...813L..28A,2016NatPh..12.1170C}.  The inner halo extends out to $\sim$10--15~kpc from the Galactic Center, has negligible net rotation, and has a median metallicity of $\text{[Fe/H]}=-1.6$~\citep{Carollo:2007xh}.  The outer halo extends beyond $\sim$15--20~kpc, exhibits retrograde motion, and has a median metallicity of $\text{[Fe/H]}=-2.2$~\citep{Carollo:2007xh}.  The outer halo is also hotter than the inner halo, with larger velocity dispersions~\citep{2010ApJ...712..692C}.  Within Solar distances of $\sim$4~kpc, the inner halo stars dominate, although there is also a sub-dominant population of outer-halo stars on highly eccentric orbits~\citep{2010ApJ...712..692C,2013ApJ...763...65A,2015ApJ...813L..28A}.  Current observational evidence supports the idea that the inner halo formed primarily from the merger of a few massive satellites, while the outer halo was built up from the disruption of many small ultrafaint dwarf galaxies~\citep{Carollo:2007xh}, although this will be further clarified with data from future \emph{Gaia} data releases.  For either of these scenarios, the DM is also stripped from the merging satellites.  For this reason, reconstructing the DM velocity distribution for direct detection experiments does not require distinguishing the inner and outer-halo stars. 
We note that our analysis does not distinguish potential contributions to the stellar halo from disrupted globular clusters~\cite{2018RSPSA.47470616F}, which should not be correlated with DM.
Further study is needed to accurately estimate the fractional contribution of globular clusters to the stellar halo.

\section{Kinematic Distribution}
\label{sec:kinematicdistribution}

%\subsection{Halo Stars}
%\label{sec:halo}

We apply the analysis strategy described in the previous section to the two separate RT samples with $\FeH<-1.5$ and $-1.8$.  The best-fit values for the parameters (16$^\text{th}$, 50$^\text{th}$, and 84$^\text{th}$  percentiles of each posterior distribution) are provided in \Tab{tab:halo}\footnote{ We found that our results are not affected by the choice of metallicities from RAVE-on \cite{2016arXiv160902914C} over those quoted in the fifth RAVE data release \cite{2017AJ....153...75K}.}, and the full triangle plots are included in the Appendix as Figs. \ref{fig:corner15} and \ref{fig:corner18}.  

\begin{figure*}[tb] %  figure placement: here, top, bottom, or page
\centering
\includegraphics[width=0.95\textwidth]{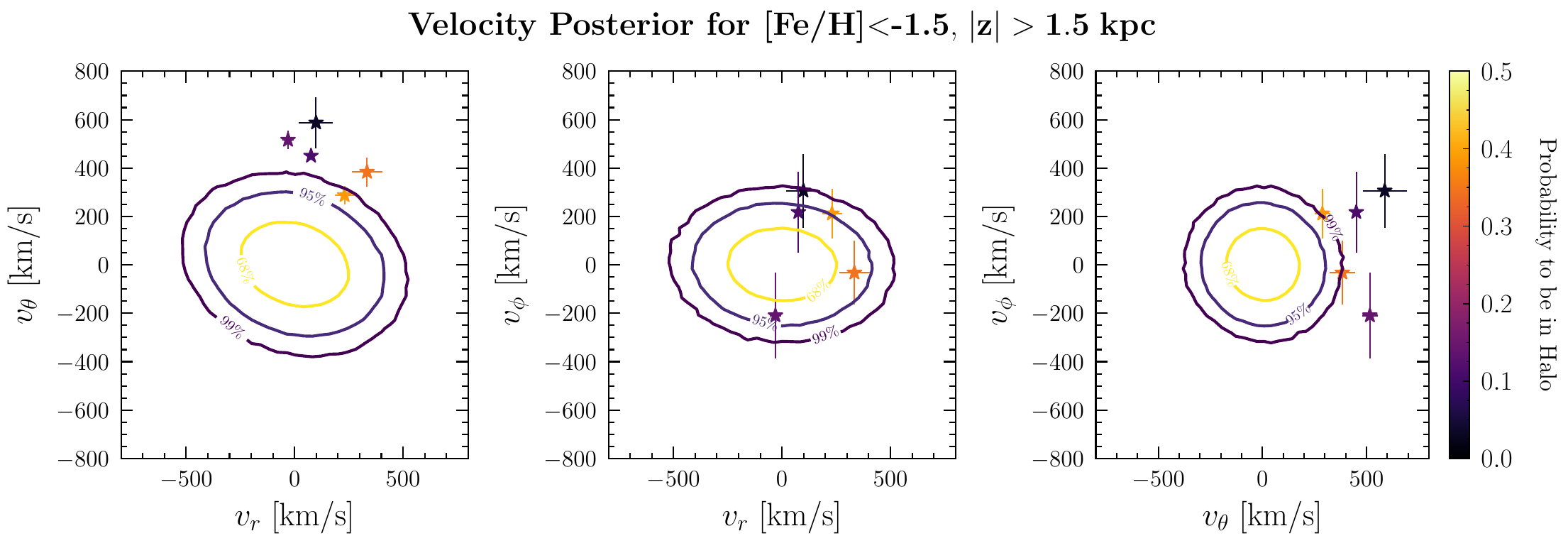} 
\includegraphics[width=0.95\textwidth]{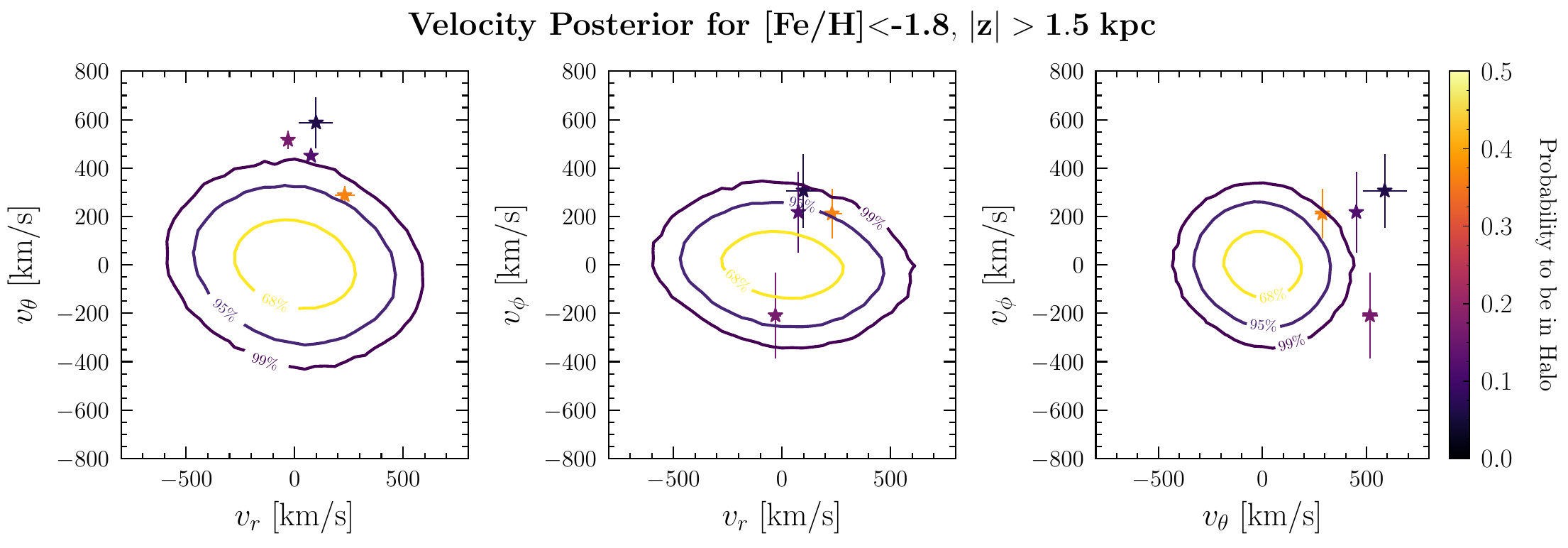} 
\caption{Posterior distributions of the spherical galactocentric velocities for the $\FeH<-1.5$ analysis (top) and $\FeH < -1.8$ analysis (bottom).  The 68\%, 95\%, and 99\% contours are shown in yellow, blue, and purple, respectively.  The stars that are identified as kinematic outliers in each analysis are also shown.  Each star's color indicates its probability of belonging to the halo, relative to the outlier, population.  The stars with the lowest probabilities tend to be located outside the 99\% halo contour. }
\label{fig:prob_plot}
\end{figure*}

The top row of Fig.~\ref{fig:prob_plot} shows the best-fit halo distributions in $v_{r,\theta,\phi}$ for the $\FeH<-1.5$ sample.\footnote{We also show these contours in \Fig{fig:prob_plot_all_stars} overlayed with all the stars in the sample. }  We find that the normal distribution for the halo stars is centered at $\mathbf{v} \sim 0$~km/s.  The only significant velocity correlation is observed between $v_r$ and $v_\theta$, where $\rho_{r\theta} = -0.18_{-0.14}^{+0.13}$.  This manifests as the tilt in the velocity ellipsoid in the top left panel of Fig.~\ref{fig:prob_plot}.  

Repeating the analysis on the $\FeH<-1.8$ sample, we find that the best-fit values for the parameters are generally consistent with those obtained using the higher-metallicity sample, except that the uncertainties are typically larger.  Again, we find that the best-fit normal distribution for the halo is centered at zero velocity.  In this case, all the correlation coefficients are consistent with zero to within their 16--84$^\text{th}$  percentiles, including $\rho_{r\theta}$ which is $-0.16_{-0.20}^{+0.20}$. 

\begin{table}[tb]
\begin{center}
\begin{tabular}{C{2.5cm}C{2.5cm}C{2.5cm}C{2.5cm}C{2.5cm}}
% \multicolumn{3}{c}{\textsc{Halo}} \\
\Xhline{3\arrayrulewidth}
\renewcommand{\arraystretch}{1.5}
Parameter & \multicolumn{2}{c}{Halo Best-Fit Values} \\%& \multicolumn{2}{c}{Outlier Best-Fit Values} \\
& $\FeH < -1.5$ & $\FeH< -1.8$ \\% $\FeH < -1.5$ & $\FeH< -1.8$ \\ 
\hline
$\mu_r$ [km/s]&  $0.48^{+5.80}_{-4.72} $ & $0.90^{+4.32}_{-3.27}$ \\%& $0.76^{+3.75}_{-3.46} $ & $0.87^{+4.16}_{-3.41} $\\
$\mu_{\theta}$ [km/s]&  $1.71^{+6.99}_{-3.98}$ & $0.79^{+4.71}_{-3.67}$ \\%&$0.82^{+5.39}_{-3.82}$ &$0.82^{+4.15}_{-3.45}$    \\
$\mu_{\phi}$ [km/s]& $ 1.53^{+5.21}_{-3.83} $ & $0.01^{+3.55}_{-4.46}$\\%&$ 0.97^{+6.14}_{-5.84} $ & $0.16^{+3.51}_{-3.35}$\\
$\sigma_r$ [km/s]& $ 164^{+15.7}_{-16.4}$ & $178^{+27.5}_{-26.1}$\\%&$ 204.79^{+115.93}_{-65.18}$ & $188.11^{+101.86}_{59.60}$\\
$\sigma_{\theta}$ [km/s]& $ 117^{+15.5}_{-16.4}$ & $121^{+27.3}_{-27.9}$\\%&$ 344.38^{+349.05}_{-121.72}$ & $299.60^{+281.21}_{-91.10}$\\
$\sigma_{\phi}$ [km/s]& $ 100^{+11.8}_{-13.1} $ & $96.5^{+22.6}_{-34.8}$\\%& $ 206.23^{+197.26}_{-66.85} $ & $205.51^{+172.19}_{-60.51}$  \\
$\rho_{r \theta}$& $-0.18^{+0.13}_{-0.14} $ & $-0.16^{+0.20}_{-0.20}$\\%& $-0.15^{+0.36}_{-0.37} $ & $0.00^{+0.37}_{-0.33}$ \\
$\rho_{r \phi}$& $-0.05^{+0.13}_{-0.13} $ & $-0.15^{+0.19}_{-0.20}$ \\%& $-0.11^{+0.44}_{-0.39} $ & $0.07^{+0.41}_{-0.36}$ \\
$\rho_{\theta \phi}$& $-0.04^{+0.14}_{-0.14}$  & $-0.10^{+0.20}_{-0.22}$ \\%& $-0.18^{+0.36}_{-0.37}$  & $0.18^{+0.33}_{-0.32}$  \\
$Q$&  $0.09^{+0.14}_{-0.06}$& $0.14^{+0.25}_{-0.09}$\\
\Xhline{3\arrayrulewidth}
\end{tabular}
\caption{\label{tab:best_fit_results} Best-fit values (16$^\text{th}$, 50$^\text{th}$, and 84$^\text{th}$ percentiles) for the halo model parameters, as well as $Q$, the average probability that a star belongs to the outlier population.}
\end{center}
\label{tab:halo}
\end{table}

%\subsection{Kinematic Outliers}
%\label{sec:kinematicoutliers}

Next, we consider the population of kinematic outliers in the RT  catalog.  Instead of discussing the average properties of this population through its velocity distribution function, we instead take the approach of identifying and studying each outlier individually.  To identify whether a star is a kinematic outlier, we calculate the probability that it belongs to the halo versus the outlier population.  To do so, we evaluate
\begin{eqnarray}
p\left(q_i \,|\, d, \theta \right) &=& \int p\left(q_i \,|\, d , \theta \right) \, p\left(\theta \,|\,d\right)\, d\theta \nonumber \\ 
&\approx& \frac{1}{N_\text{mc}} \sum_{n=1}^{N_\text{mc}} p\left( q_i \,\Big|\, d, \theta^{(n)} \right) \, ,
\label{eq:qi}
\end{eqnarray}
where $d$ is shorthand for the full data set containing the velocities and errors for each star (\emph{i.e.}, $\{\mathbf{v}_i\}, \{\boldsymbol{\sigma}_i\}$).  Here, $\theta^{(n)}$ denotes the set of parameters sampled in the $n^\text{th}$ position of the Markov chain of length $N_\text{mc}$.  

Eq.~\ref{eq:qi} allows us to recover the probability that a star in the RT sample belongs to the halo.  We label star $i$ as an outlier if 
\begin{equation}\label{eq:outlier}
p\left(q_i = 0\,|\, d, \theta \right)  \leq 0.5 \quad \text{(outlier condition)}  \,.
\end{equation}
That is, if it has a probability of less than 50\% to belong to the halo.  In general, we find that the stars break up cleanly into two populations above and below the 50\% cutoff (see \Fig{fig:prob_distribution}).  The values of all chemical, spatial and kinematic properties of  the outliers are listed in \Tab{tab:stars}.

We identify five (four) stars as outliers in the [Fe/H]$<-1.5$ ($-1.8$) samples. Four of these stars are common to both;  one only appears in the $\FeH<-1.5$ analysis because its iron abundance is $\FeH=-1.69$.  The five outlier stars are giants, with surface gravities in the range $0.09 \lesssim \log g \lesssim 0.95$~dex and effective temperatures $4360 \lesssim T_\text{eff} \lesssim 4730$~K.  
These stars are metal-poor with iron abundances from $-2.56\lesssim$$\FeH\lesssim-1.69$, as determined by RAVE-on.  Additionally, they fall within 3.6--6.4~kpc of the Sun, and exhibit no obvious spatial clustering, as illustrated in the left panel of Fig.~\ref{fig:zRgc}.  One might note that two of these stars appear to be relatively close to each other.  However, we find that their orbits are largely separated, and therefore do not believe that they share a common origin.

The outliers are indicated as individual points in the panels of Fig.~\ref{fig:prob_plot}, where the color of each point corresponds to its probability of being a halo star, rather than an outlier.  For the most part, the stars lie outside the 95\% velocity contours for the halo component, with many lying outside the 99\% contour.  In general, the stars are distributed symmetrically around $v_{r, \phi} \sim 0$, but are skewed towards positive $v_\theta$.  
 
\begin{figure*}[t] %  figure placement: here, top, bottom, or page
\centering
\includegraphics[width=0.45\textwidth]{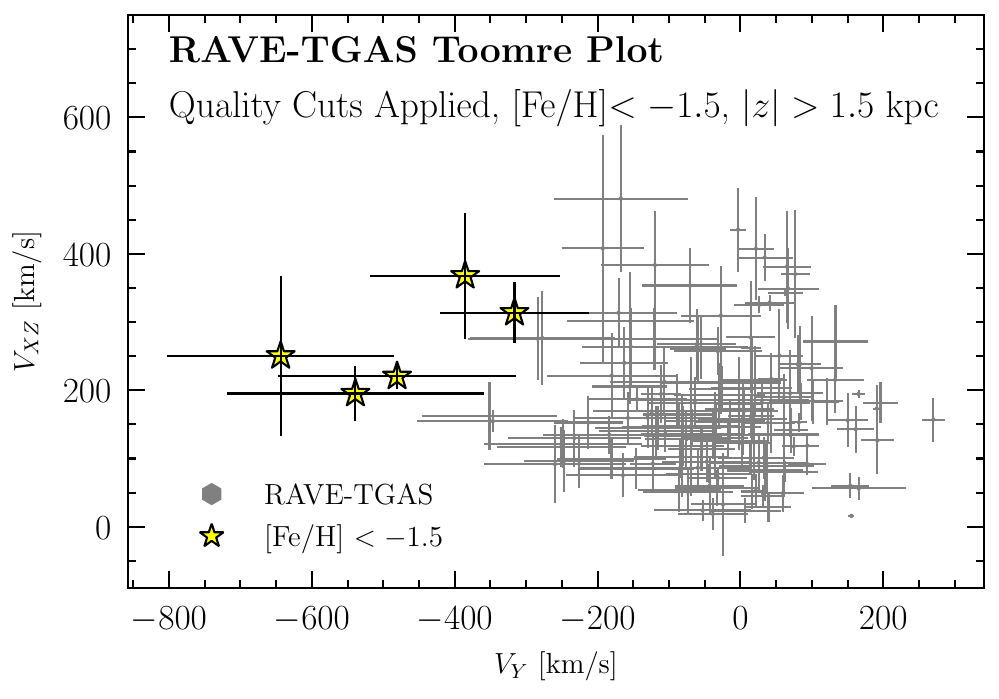} 
   \qquad
\includegraphics[width=0.45\textwidth]{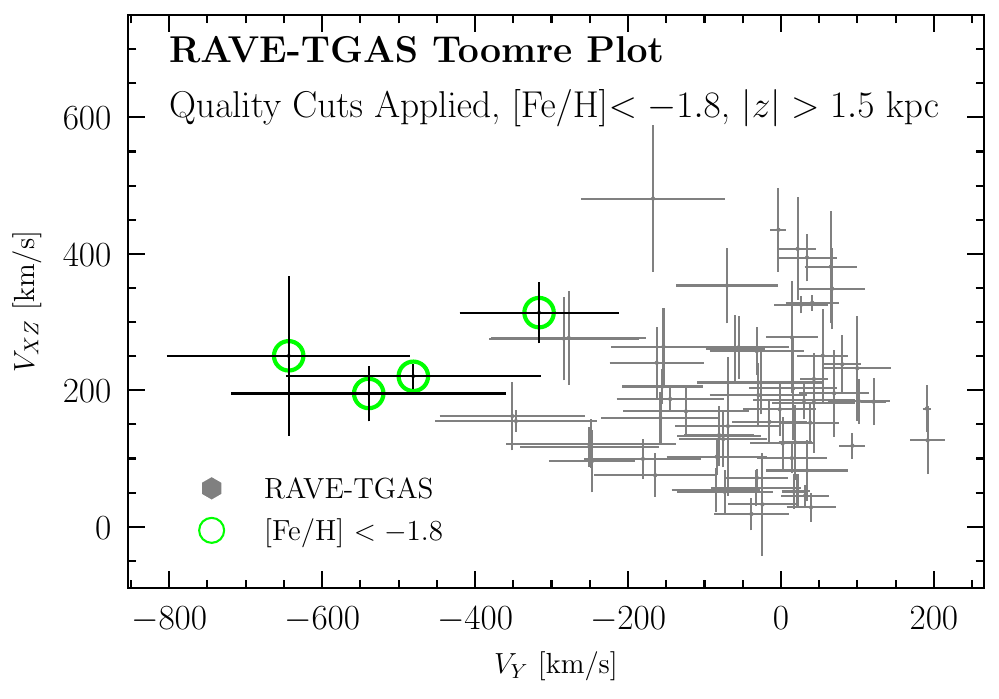} 
\caption{Toomre plot of all stars in the RAVE-TGAS sample with $\text{[Fe/H]}<-1.5~(-1.8)$ in the left~(right) panel.  The kinematic outliers for each sample are indicated by the yellow stars~(green circles) in the left~(right) panel.}
\label{fig:energy_lz}
\end{figure*}

 Figure~\ref{fig:energy_lz} shows the distribution of velocities in the $V_{XZ}$--$V_Y$ space, where $V_{XZ}=\sqrt{V_X^2 + V_Z^2}$ in the galactocentric cartesian frame for the $\text{[Fe/H]}<-1.5~(-1.8)$ sample on the left~(right) panel.  The stars identified as belonging to the halo component are shown in gray.  The error bars are computed by propagating the errors on the distances, proper motions, and heliocentric velocities \citep{Bovy:2011aa}.  The outliers are identified as yellow stars~(green circles) for the $\FeH < -1.5$~($-1.8$) sample.  While the estimated velocity errors for these stars are quite large, we do find that they have high velocities.  Additionally, they appear to be clustered at negative $V_Y$ and exhibit retrograde motion.  

At this stage, the origin of these outliers is not known.  One likely possibility is that they are outer-halo stars on eccentric orbits just passing through the local volume; this population is expected to be very metal-poor and exhibit retrograde motion~\citep{2010ApJ...712..692C}.  Another intriguing possibility is that the outliers might be tidal debris associated with the disruption of a nearby satellite.  In this case, the halo population recovered by the statistical analysis would be a mixture of both inner and outer-halo stars, and the outliers would be associated with kinematic substructure.                                                  

We have searched for references of these outlier stars in the literature.  There is no overlap between our outliers and those identified in a separate study of the RT data by~\cite{2016arXiv161100222H}.  Little information appears to be known for four of the outliers we identify, beyond their chemical and kinematic properties from RAVE and---in some cases---an entry in the Ecliptic Plane Input Catalog~\citep{2011AJ....141..187S,2014AJ....148...81M,2016ApJS..224....2H}.  However, the star with ID `20070811\_1523m09\_120' (also referred to as  HE 1523-0901) is a very r-process enhanced metal-poor star discovered by \cite{2007ApJ...660L.117F}. It would be interesting to trace the orbital origin of this star, in particular from other r-process dwarfs such as Reticulum II \citep{2015ApJ...805..130K,2016ApJ...830...93J} and Tucana III \citep{2017ApJ...838...11S,2017ApJ...838...44H}.  

If the origin of the kinematic outliers can be traced to dwarf galaxies, it may suggest evidence for correlated substructure in the local DM phase space, which would have significant ramifications for direct detection experiments.  Improved measurements of the parallaxes and proper motions for these stars will reduce the errors on their derived velocities, while dedicated spectral studies can refine estimates of their chemical composition.  In future work, we plan to explore the orbital properties of these outliers to better understand whether they can be traced to known dwarfs, or whether their maximum distance off the plane and/or eccentricity are more consistent with the expectation of outer-halo stars.

\section{Dark Matter Implications} 
\label{sec:DM}

 \begin{figure*}[t] %  figure placement: here, top, bottom, or page
   \centering
   \includegraphics[width=0.45\textwidth]{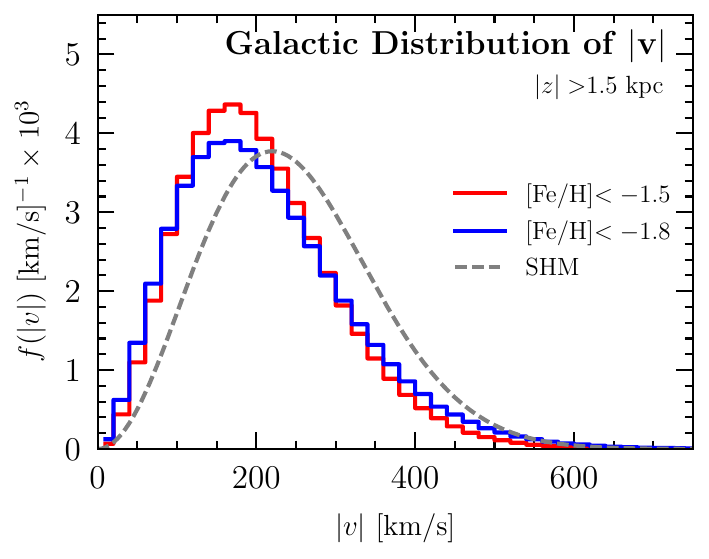} 
   \qquad
   \includegraphics[width=0.45\textwidth]{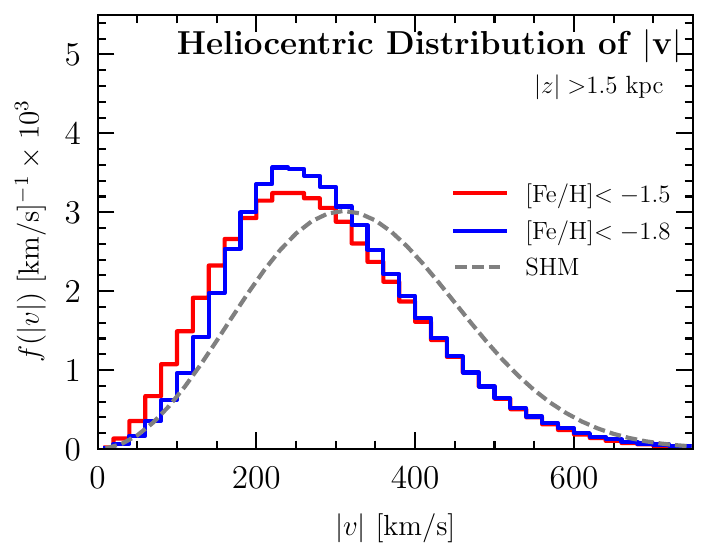} 
   \caption{The posterior speed distribution for the metal-poor stars in the RAVE-TGAS halo, derived from the best-fit parameters in Tab.~\ref{tab:halo}.  The results for the $\FeH < -1.5$ and $-1.8$ samples are shown separately in red and blue, respectively.  The gray-dashed line represents the Standard Halo Model (SHM).  The distributions are shown for the Galactic (left) and heliocentric (right) frames.   Using the full nine-dimensional posterior distribution (after marginalizing over the outlier parameters), we find that the SHM is $\sim6\sigma$ discrepant with the empirical distributions.  Interpolations of the heliocentric distributions are provided as Supplementary Data.   } \label{fig:vel_distribution}
\end{figure*}

In this section, we assume that the best-fit RT distributions for the metal-poor halo obtained above are adequate tracers for the kinematics of the virialized DM, and explore the implications for direct detection experiments.  The left panel of \Fig{fig:vel_distribution} shows the posterior distributions for the stellar speeds in the Galactocentric frame for the $\FeH <-1.5$ and $-1.8$ samples (red and blue lines, respectively).  The two distributions are both peaked at $\sim$$170$~km/s and are consistent with each other.  The more metal-poor sample has a slightly larger dispersion, as we expect from the results presented in Sec.~\ref{sec:kinematicdistribution}.  For comparison, we also show the Standard Halo Model (SHM), defined as 
\begin{equation}\label{eq:maxwell}
f_\text{SHM}(v)=\frac{4 \, v^2}{\sqrt{\pi} \, v_c^3} \, \exp\left[ -\frac{v^2}{v_c^2}\right]\, ,
\end{equation} 
where $v_c \sim 220$~km/s is the local circular velocity~\citep{Kerr:1986hz,Reid:2009nj,McMillan:2009yr}.  We see that the RT samples have a lower average speed 
than expected from the SHM.  Additionally, the RT distributions predict fewer high-speed particles above $v\sim200$~km/s.  

\Fig{fig:vel_distribution} shows that there is a large discrepancy between the velocity distributions of the metal-poor stars and the SHM. To quantify this difference, we marginalize over the kinematic outlier parameters in the MCMC, and integrate the resulting nine-dimensional posterior distribution for the halo parameters to find the corresponding p-value associated with the SHM.  We find that the empirical distribution is $\sim$$6\sigma$ discrepant with the SHM,\footnote{Specifically, the  probability of obtaining the SHM values from the full posterior distribution is $\sim$$10^{-8}$.} where the discrepancy is primarily due to the fact that the observed distribution is anisotropic.

While our analysis method accounts for uncertainties on the measured distances and proper motions of the stars in the RT sample, it does not account for systematic biases in these measurements.  As already mentioned,~\cite{2017arXiv170704554M} showed that TGAS parallaxes for stars with $\log g \lesssim 2.0$ are systematically higher than those from RAVE DR5, which could over-predict the distances and, thus, the stellar velocities.  About $93\%$~($91\%$) of the  stars in the $\FeH<-1.5$~($-1.8$) analysis fall in this range, and it is difficult to assess the impact on the results of our analysis.  These challenges will be ameliorated, however, as the data continues to improve with future \emph{Gaia} data releases.

 In the meantime, however, it is worthwhile understanding how the  metal-poor RT speed distributions shown in \Fig{fig:vel_distribution} could impact the scattering rates in direct detection experiments, if we assume that they effectively trace the DM distribution.  To do so, we need to transform the speeds to the heliocentric frame.  The results are shown in the right panel of \Fig{fig:vel_distribution}.\footnote{When transforming the RT posterior speed distribution to the heliocentric frame, we use the observed angular distributions of the stellar sample.  The results are essentially identical if we instead assume a uniform angular distribution.}  We have found that neither the Maxwell-Boltzmann distribution nor the function proposed by \cite{Mao:2012hf}  provide a satisfactory fit, and  we therefore provide interpolations of the heliocentric RT distributions  as tables in the Supplementary Data.\footnote{\label{link} \url{https://linoush.github.io/DM_Velocity_Distribution/}} 

The rate for a DM particle of mass $m_\chi$ to scatter off a nucleus of mass $m_N$ is proportional to 
\begin{equation}
g(v_\text{min})=\int_{v_\text{min}}^{\infty} \frac{\tilde{f}(v)}{v}\, dv \, , 
\label{eq:gvmin}
\end{equation}
where $\tilde{f}(v)$ is the speed distribution in the lab frame and $v_\text{min}$ is the minimum DM speed needed to produce a nuclear recoil of energy $E_\text{nr}$~\citep{1996PhR...267..195J,Freese:2012xd}.
Assumptions about the particle physics model for the DM candidate factor into $v_\text{min}$. For elastic scattering,
\begin{equation} \label{eq:vmin}
v_\text{min} =\sqrt{\frac{E_\text{nr} \, m_N}{2 \, \mu^2}} \, ,
\end{equation}
where $\mu = m_\chi m_N/ (m_\chi + m_N)$ is the reduced mass.  For a specific nuclear target, $v_\text{min}$ is larger for experiments with higher energy thresholds and/or lower DM mass.  In either case, $g(v_\text{min})$ becomes increasingly more sensitive to the high-speed tail of the $\tilde{f}(v)$ distribution.  

\begin{figure}[t] %  figure placement: here, top, bottom, or page
   \centering
   \includegraphics[width=0.45\textwidth]{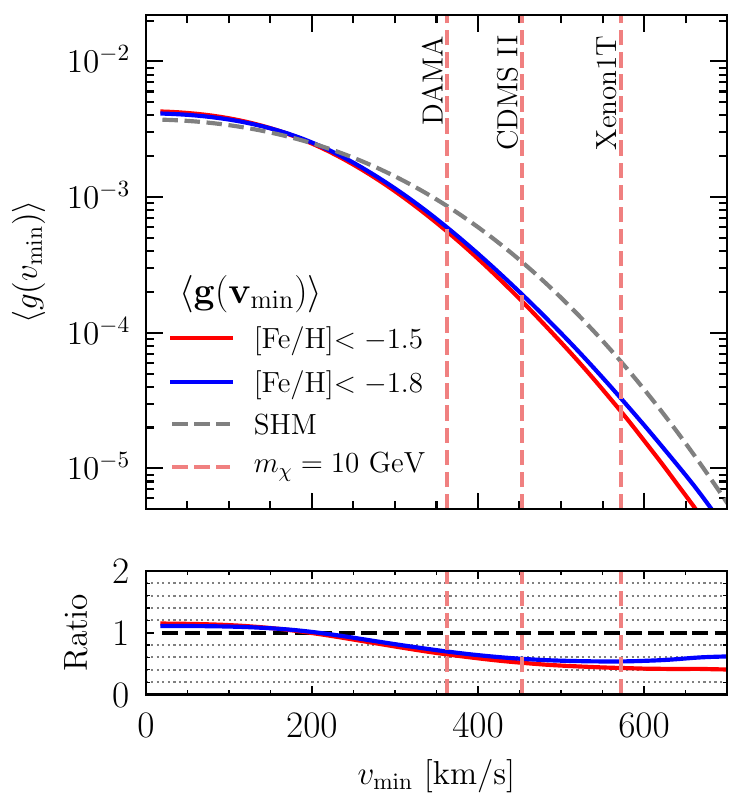} 
   \caption{The function $\langle g(v_{\text{min}}) \rangle$, defined in \Eq{eq:gvmin} and averaged over a year, plotted in terms of the minimum scattering velocity, $v_{\text{min}}$, for the $\FeH < -1.5$ and $-1.8$ RAVE-TGAS samples (red and blue, respectively). For comparison, we also show $\langle g(v_{\text{min}}) \rangle$ for the SHM in dashed gray. The ratio between the RAVE-TGAS expectation and the SHM is included in the bottom panel for both iron abundances.  In the top panel, we overlay the value of $v_{\text{min}}$ that corresponds to a 10~GeV dark matter particle scattering at threshold for the DAMA/LIBRA~\citep{Bernabei:2008yi}, Xenon1T~\cite{Aprile:2017ngb}, and CDMS~II~\citep{Agnese:2017jvy} experiments. } \label{fig:gvmin}
\end{figure}

Figure~\ref{fig:gvmin} shows $\langle g(v_\text{min}) \rangle$ for both the RT distributions and the SHM, where the brackets indicate the yearly average.  Compared to the SHM, the scattering rate is larger by $\sim$$10\%$ for the RT distributions below $v_\text{min}\sim 200$~km/s, but suppressed by $\sim$$40$--$60\%$ at larger velocities.  A given direct detection experiment only probes a particular range of $v_\text{min}$, dependent on its nuclear target and the energy range that it is sensitive to. Consider, for example, the Xenon1T experiment~\citep{Aprile:2017iyp}, which uses a xenon target to search for nuclear recoils in the range $E_\text{nr} = \text{[5, 40]}$~keV.  For $m_\chi = 50$~GeV, this corresponds to $v_\text{min} \sim 150$~km/s at threshold (\emph{e.g.}, $E_\text{nr}=5$~keV).  In this case, the ratio of $\langle g(v_\text{min}) \rangle$ for the RT and SHM distributions is close to unity, and the scattering rates are essentially equivalent for either velocity distribution.  In contrast, when $m_\chi = 10$~GeV, then the same experiment is sensitive to $v_\text{min} \sim 570$~km/s at threshold.  At this minimum speed, the RT distribution results in a scattering rate that is a factor of $\sim$$2$ smaller than that obtained with the SHM.

The dashed vertical lines in \Fig{fig:gvmin} indicate the value of $v_\text{min}$ that each experiment is sensitive to for a $m_\chi =10$~GeV DM particle scattering at threshold.  To compare with the Xenon1T example described above, we also show the results for the DAMA/LIBRA~\citep{Bernabei:2008yi} and CDMS~II~\citep{Agnese:2015ywx} experiments.  For DAMA, a 10~GeV particle preferentially scatters off the sodium in the crystal lattice; at the threshold energy of $\sim$$6.7$~keV, $v_\text{min} \sim 350$~km/s.  For CDMS~II, which uses a germanium target, $v_\text{min}\sim450$~km/s for a $\sim$$5$~keV threshold. In all these cases, the expected scattering rate is a factor of 2 too strong when the SHM is used rather than the empirical velocity distribution.  These discrepancies are particularly relevant for experiments that are aimed at probing DM masses below $\sim$GeV.
For example, CDMSlite~\citep{Agnese:2017jvy} has a low energy threshold and can probe DM masses below $\sim$$3$~GeV, where it is sensitive to the high-speed tail of the DM distribution. 

We stress that the arguments presented in this section apply to the virialized DM halo. 
If the kinematic outliers are associated with local DM substructure---such as streams or debris flow~\citep{2008Natur.454..735D,2011MNRAS.413.1419V,Freese:2003na,Savage:2006qr,Kuhlen:2009vh,Lisanti:2011as,Kuhlen:2012fz,Lisanti:2014dva}---they could also impact the interpretation of the experimental results.  Because the origin of these outliers is unknown, however, we choose to not include them when evaluating  $\langle g(v_\text{min})\rangle$.  Additionally, recent mergers of satellite galaxies could lead to spatial or kinematic substructure that is associated with more metal-rich stellar populations in the halo above $\FeH \sim -1.5$.  The analysis we used here cannot simply be extended to higher metallicities, where one must  contend with increased contamination from disk and \emph{in-situ} halo stars.  However, it would be worthwhile to consider ways of generalizing the statistical methods of \Sec{sec:stats} to explore stellar substructure in this regime.    

\section{Conclusions} \label{sec:conclusions}

The primary goal of this paper was to infer the local distribution of virialized DM velocities using metal-poor stars as kinematic tracers.  To achieve this, we took advantage of the RAVE-TGAS catalog, which currently provides one of the best maps of the local phase-space of the stellar halo.  We conservatively required that all stars be located $|z|>1.5$~kpc from the Galactic plane to reduce potential contamination from the disk and \emph{in-situ} stars, and considered two samples of low-metallicity stars: $\FeH<-1.5$ and $-1.8$.  This allowed us to probe the  population of \emph{ex-situ} halo stars, while still maintaining a sample size large enough for the statistical methods to converge.   

To recover the velocity distribution for these stars, we used a Gaussian mixture model that accounted for both a stellar halo component and a population of kinematic outliers.  The best-fit model parameters for the halo population were consistent, within uncertainties, between the $\FeH<-1.5$ and $-1.8$ samples, despite the fact that the latter includes nearly half the number of stars as the former. We can compare the best-fit speed distributions that we recover for the halo component with results from  previous work using the Sloan Digital Sky Survey.  In particular, \cite{2010ApJ...712..692C} provide the velocity components for stars within 4~kpc of the Sun and Galactocentric radii $7 < r < 10$~kpc.  We can estimate the corresponding speed distributions from the distributions they provide in the metallicity bins $\FeH \in [-2,-1.6]$ and $[-2.2, -2]$, if we make the simplifying assumption that the velocities are uncorrelated.  Our results are in good agreement with theirs in this range.  Due to the limited sample size of the RT dataset and the fact that it is not volume complete, we cannot extract the density distribution of the population of stars studied here.  However, \cite{2010ApJ...712..692C} showed that the density of the most metal-poor stars ($\FeH \sim -2.2$) in their sample is proportional to $\sim r^{ -2}$, similar to the isothermal density distribution expected of DM.    This density distribution is derived kinematically using Jeans equation and assuming that the system is in dynamic steady state.

 The RT sample quickly becomes statistics-limited below $\FeH \lesssim -2$.  As a result, we focused on the distribution for $\FeH <-1.5$ and $-1.8$.  In the previous SDSS study of the local halo~\citep{2010ApJ...712..692C}, the velocity distribution was probed down to $\FeH < -2.2$.  They found that the dispersion of the resulting speed distribution increased because the sample was dominated by outer-halo stars, which are generally hotter than inner-halo stars.  For the purpose of reconstructing the DM velocity distribution, it will be imperative to verify these conclusions---and continue characterizing the kinematics at even lower metallicities---with the upcoming \emph{Gaia} data release. 

The posterior speed distributions of the metal-poor halo stars were obtained from the best-fit multivariate distribution.  Assuming this distribution matched that of the DM,  a result substantiated by numerical simulations \cite{eris_paper}, we studied the implications for direct detection experiments.  Our point of comparison was the SHM, which is the most common distribution that is assumed for the DM velocities.  The best-fit dispersions that we recovered are $\sim$$6\sigma$ discrepant with the SHM values. The discrepancy mainly arises because the distribution we recover is not isotropic, in contrast to the SHM.  For DM masses below $\sim$$10$~GeV, the SHM typically predicts a scattering rate that is nearly a factor of two larger than what is expected from the empirical distribution.  

In addition to characterizing the halo population, our analysis also identified velocity outliers.  
Improved kinematic and spectral studies will help determine whether these stars point to local phase-space substructure that is correlated with tidal debris from a satellite merger.  If so, then the kinematic substructure in the stellar distribution may also be correlated with DM.  Because we cannot make any firm conclusions at this stage, we did not consider the implications of these outliers for direct detection.

To summarize, we found the best-fit velocity distribution of the local metal-poor stellar halo using data from RAVE-TGAS and used these results to motivate an empirical distribution for the virialized dark matter component. \emph{Gaia} is anticipated to revolutionize our understanding of the phase-space distribution of the stellar halo.  As the uncertainties on the stellar parameters continue to improve for an increasingly larger selection of stars, we expect to further refine our characterization of the smooth dark matter distribution and any potential substructure.  

For the interested reader, we provide the posterior speed distributions for the $\FeH < -1.5$ and $-1.8$ samples in the heliocentric frame as Supplementary Data.\footnoteref{link}

\section*{Acknowledgements}
We thank S.~Agarwalla, A.~Bonaca, A.~Frebel, D.~Hogg, A.~Ji, K.~Johnston, A.~Leder, P.~Madau,  A.~Peter, A.~Price-Whelan, D.~Spergel, and J.~Wojno for helpful conversations. M.L. is supported by the DOE under contract DESC0007968, the Alfred P.~Sloan Foundation, and the Cottrell Scholar program through the Research Corporation for Science Advancement.  L.N. is supported by the DOE under contract DESC00012567.

%\clearpage
%\def\bibsection{} 

\newpage
\appendix
\section{Supplemental Discussion}

\setcounter{equation}{0}
\setcounter{figure}{0}
\setcounter{table}{0}
\setcounter{section}{0}
\makeatletter
\renewcommand{\theequation}{S\arabic{equation}}
\renewcommand{\thefigure}{S\arabic{figure}}
\renewcommand{\thetable}{S\arabic{table}}

\begin{figure*}[b] %  figure placement: here, top, bottom, or page
   \centering
   \includegraphics[width=0.95\textwidth]{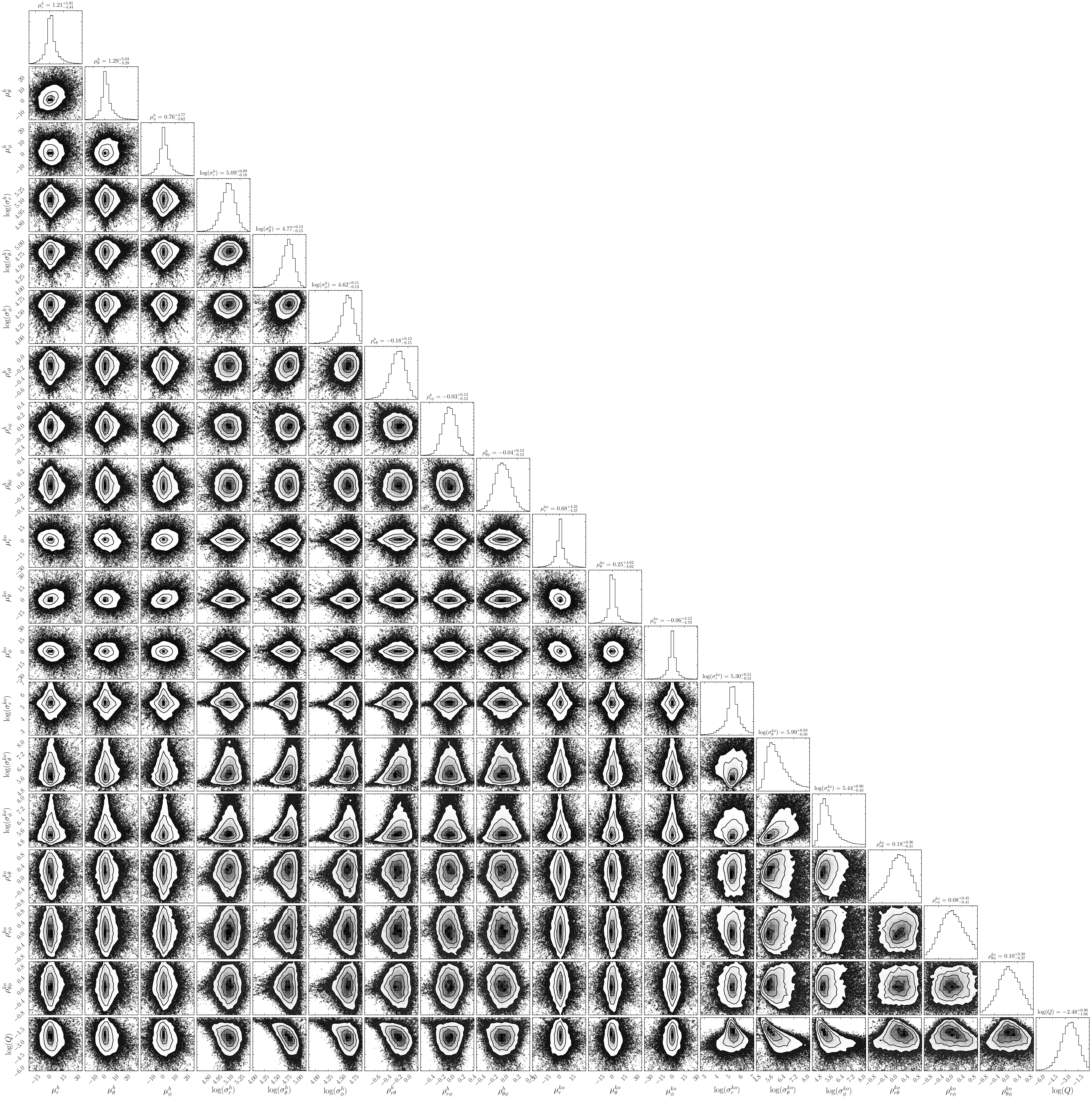} 
   \caption{Corner plot for the $\FeH<-1.5$ analysis. See \Sec{sec:stats} for more details.}
  \label{fig:corner15}
\end{figure*}

\begin{figure*}[b] %  figure placement: here, top, bottom, or page
   \centering
   \includegraphics[width=0.95\textwidth]{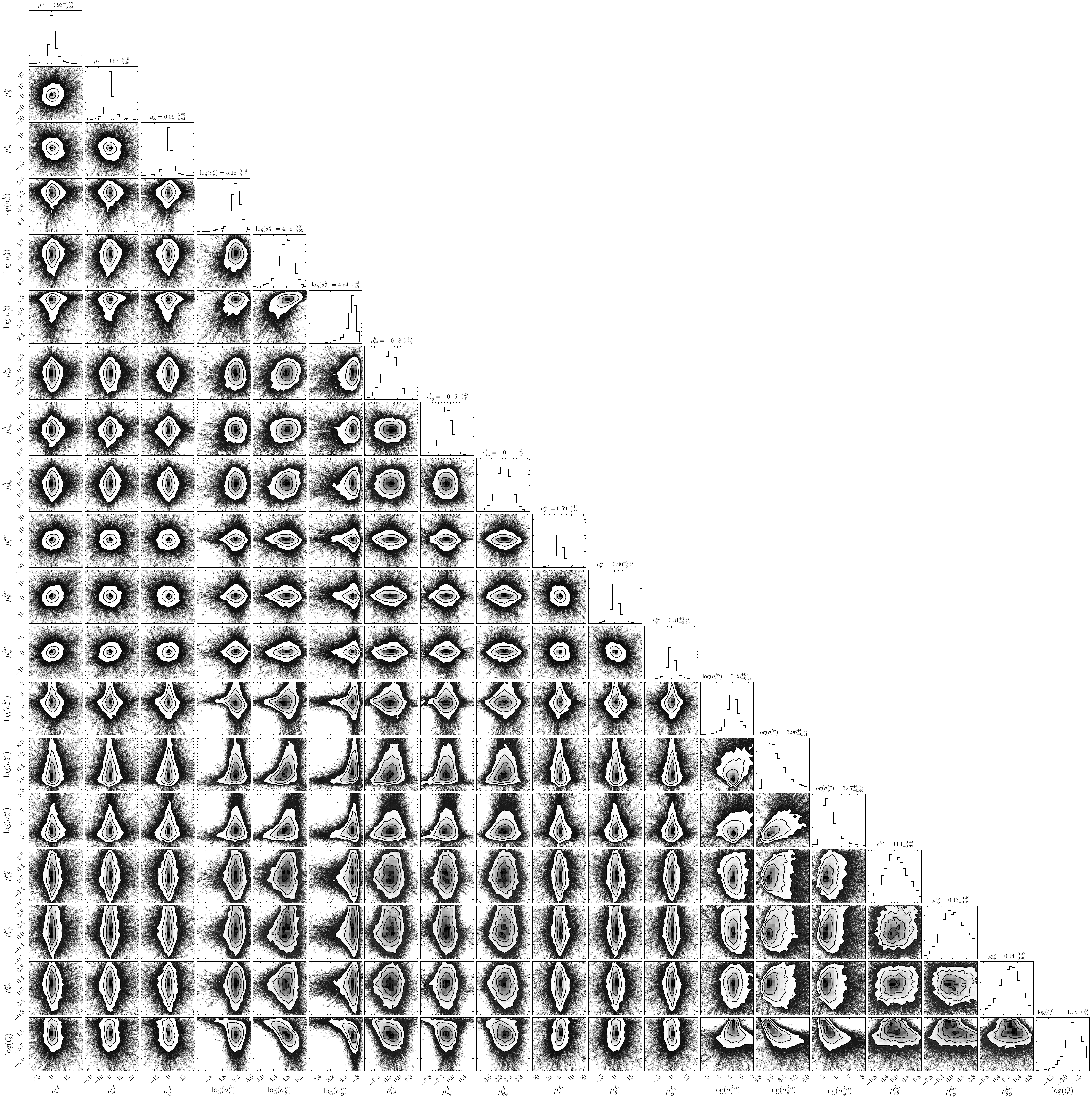} 
   \caption{Corner plot for the $\FeH<-1.8$ analysis.  See \Sec{sec:stats} for more details. }
  \label{fig:corner18}
\end{figure*}

\begin{figure*}[tb] %  figure placement: here, top, bottom, or page
\centering
\includegraphics[width=0.95\textwidth]{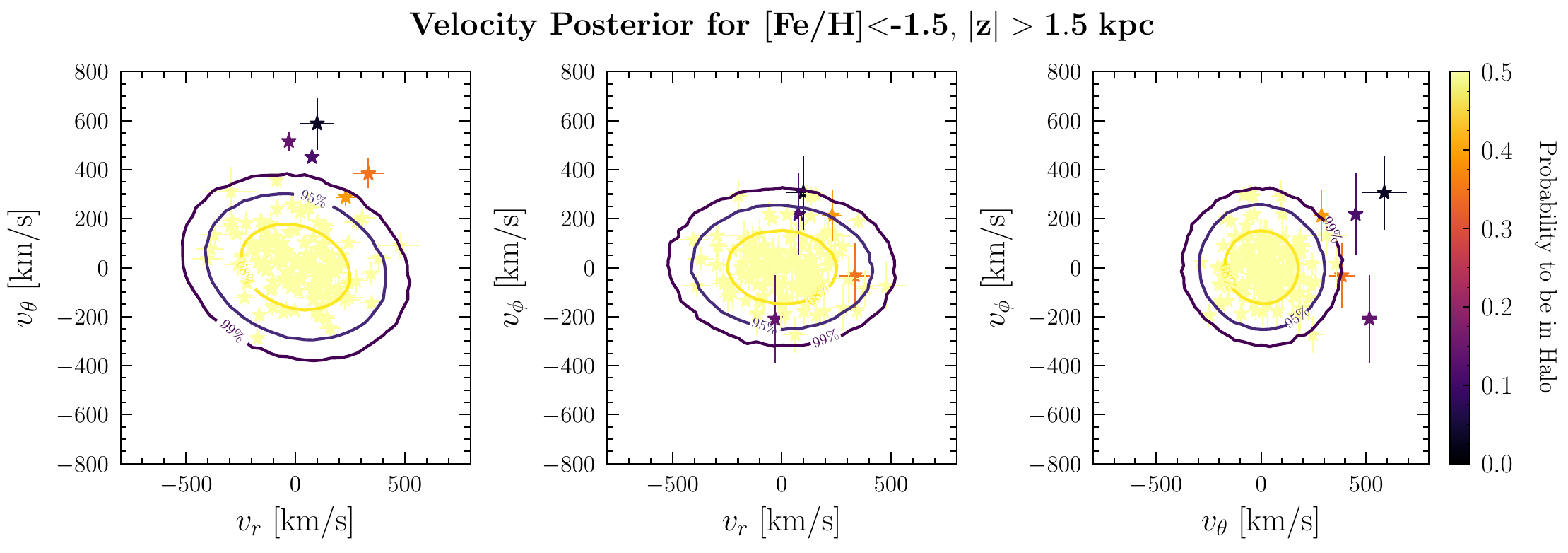} 
\includegraphics[width=0.95\textwidth]{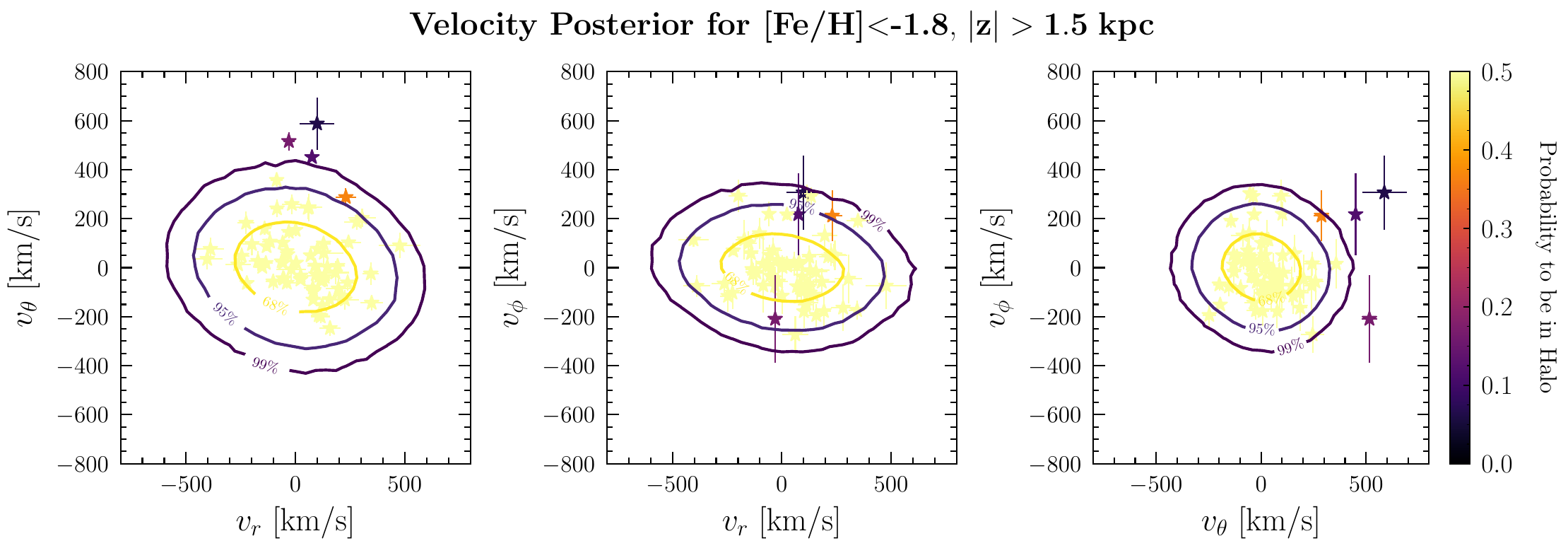} 
\caption{The same as \Fig{fig:prob_plot} in the main text, except that we now show all stars that pass the metallicity and $|z|>1.5$~kpc cuts.  Each star's color indicates its probability of belonging to the halo; note that all stars with probability $>0.5$ are indicated in yellow.  The stars with the lowest probabilities tend to be located outside the 99\% halo contour. }
\label{fig:prob_plot_all_stars}
\end{figure*}

\begin{figure*}[h] %  figure placement: here, top, bottom, or page
   \centering
   \includegraphics[width=0.44\textwidth]{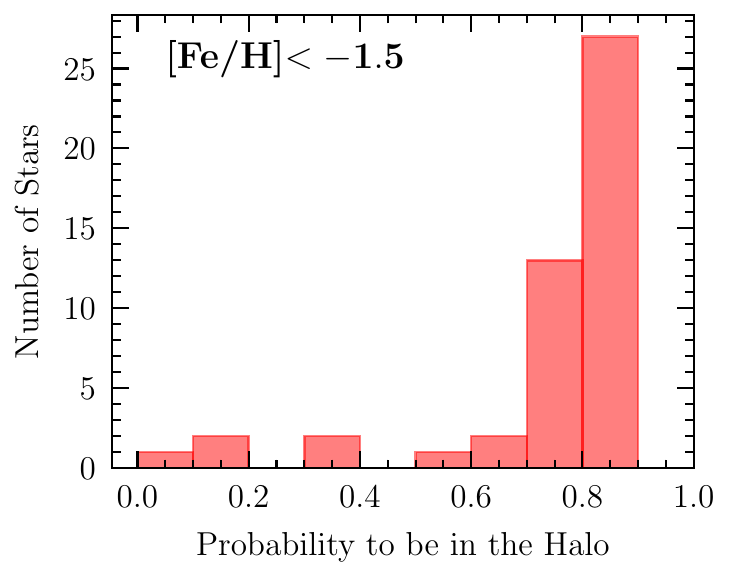} 
   \qquad
   \includegraphics[width=0.43\textwidth]{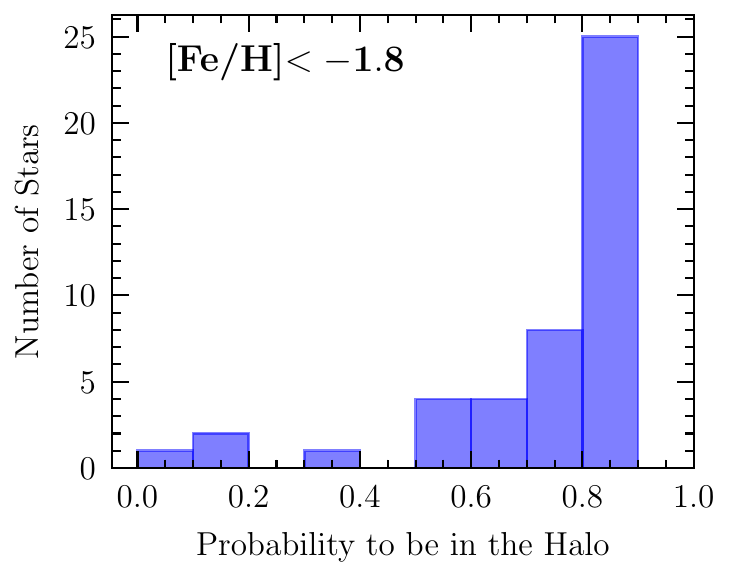} 
   \caption{Probability of the stars to belong to the halo, relative to the outlier population, for the $\FeH<-1.5$ (left) and $\FeH<-1.8$ (right) analyses. We note the presence of two separate populations over and below the $50\%$ cutoff. } \label{fig:prob_distribution}
\end{figure*}

This Appendix includes some additional figures that supplement the discussion in the main text.  \Fig{fig:corner15} and \Fig{fig:corner18} show the full triangle plots for the mixture model analysis performed on the RAVE-TGAS sample.  \Fig{fig:prob_plot_all_stars} shows the three-dimensional velocity contours overlaid with all the stars that were analyzed.  \Fig{fig:prob_distribution} histograms the stars' probabilities of belonging to the halo relative to the outlier population.  \Tab{tab:stars} lists the kinematic outliers and their properties.

\clearpage
\newpage

\clearpage
\newpage

\afterpage{%
    \thispagestyle{empty}% empty page style (?)
\begin{sidewaystable*}[h] %[p]
\centering
\begin{tabular}{lrrrrrrrrrrrrr}
\hline
 TGAS ID              &   $\alpha$ &   $\delta$ &   [Fe/H] &   $v_r$ &   $v_\theta$  &  $v_\phi$ & $\mu_{RA}$     & $\mu_{DE}$     & HRV   &   $\log(g)$ &   T$_{\text{eff}}$  & G-band & Distance\\
         &    ($^\circ$) &    ($^\circ$) &   (dex) &    (km/s) &   (km/s) &    (km/s) & (mas)    & (mas)     &  (km/s)   &   &    ($10^3$K) & (Mag) & (kpc)\\
\hline
$^{*\dagger}$ 20070811\_1523m09\_120 &       231.504 &       -9.19408 &          -2.56 &   $75.7^{\pm 15.0}$ &   $451^{\pm  166}$ &     $218^{\pm 16.1}$      &           $-22.6^{\pm 0.8}$ &       $-28.0^{\pm 0.4}$  &    $-165^{\pm 1.19}$  &     0.09 &     $ 4.36$ &     10.6 &            $4.2^{\pm 1.0}$ \\

$^{*\dagger}$ 20071017\_2109m51\_072 &     316.494 &      -49.326   &          -2.13 &    $231^{\pm   46.0}$ &    $288^{\pm  104}$ &      $ 212^{\pm   37.2}$      &         $14.2 ^{\pm     0.5}$ &      $ -26.9 ^{\pm     1.1}$  &     $213^{\pm     1.0}$ &     0.58 &      $4.49$ &     10.5 &            $4.1^{\pm      0.8}$ \\

$^{*}$  20080919\_2308m32\_143 &    347.004 &      -33.6352   &          -1.69 &    $334^{\pm    70.0}$ &     $ 384^{\pm  132}$ &      $ -32.2^{\pm  60.7} $     &    $ 22.4^{\pm      1.8 }$ &       $-18.7^{\pm 1.2}$  &    $10.1^{\pm  1.1}$  &     0.34 &      $4.6$  &     11.4 &            $5.1^{  \pm     1.1 }$ \\

$^{*\dagger}$ 20090812\_2213m50\_037 &       328.962 &      -51.0937  &          -1.97 & $ -30.2^{\pm   25.3}$ &    $516^{\pm  178}$ &  $   -209^{\pm   37.5}$      &          $ -0.26 ^{\pm      0.40 }$ &      $ -45.3 ^{\pm      0.95}$ &  $ -31.8   ^{\pm  0.55}$  &     0.95 &      $4.59$ &     11.4 &          $  3.6 ^{\pm      0.9}$ \\

 $^{*\dagger}$ 20100124\_1148m06\_070 &      176.17  &       -4.16422  &          -2.09 &   $99.0^{\pm    78.5}$ &    $587^{\pm  152}$ &     $  306 ^{\pm   106}$ &        $-21.6 ^{\pm     2.38}$  &     $  -34.6 ^{\pm      0.9}$ &    $ 292^{\pm     1.0}$ &     0.95 &      $4.73$ &     10.7 &           $ 4.4 ^{\pm       1.0 }$  \\

\hline
\end{tabular}
        \captionof{table}{List of kinematic outliers found in the analysis of \Sec{sec:kinematicdistribution}. Stars with the sign $^*$ have been found in the analysis with $\FeH<-1.5$, while those with the symbol $^\dagger$ are found in the $\FeH<-1.8$ study. } \label{tab:stars}
        \end{sidewaystable*}
}    \clearpage% Flush page

\bibliographystyle{JHEP}
\bibliography{ravetgas}

\end{document}